\documentclass[aps,prl,twocolumn,superscriptaddress,showpacs,amsmath,amssymb,longbibliography]{revtex4-2}
\usepackage[english]{babel}
\usepackage{amsmath}
\usepackage{bm}
\usepackage{graphicx,bbm} 
\usepackage{times}
\usepackage{epsfig} 
\usepackage{soul}
\usepackage[utf8]{inputenc}
\usepackage[colorlinks,linkcolor=blue,citecolor=blue,urlcolor=blue]{hyperref}
\usepackage{color}
\usepackage{latexsym}
\usepackage{ulem}
\definecolor{nred} {RGB}{224,0,0}
\definecolor{nblue} {RGB}{28,130,185}
\definecolor{dgreen} {RGB}{78,138,21}
\definecolor{norange}{RGB}{230,120,20}

\begin{document}

\title{
%1. Probing quantum phase transitions with single-qubit decoherence\\
Harnessing spin-qubit decoherence to probe strongly-interacting quantum systems
%Impurity relaxation within a strongly-correlated system
}

\author{M. Płodzień}
\affiliation{ICFO-Institut de Ciencies Fotoniques, The Barcelona Institute of Science and Technology, 08860 Castelldefels (Barcelona), Spain}

\author{S. Das}
\affiliation{FZU - Institute of Physics of the Czech Academy of Sciences, 182 00 Prague, Czech Republic}

\author{M. Lewenstein}
\affiliation{ICFO-Institut de Ciencies Fotoniques, The Barcelona Institute of Science and Technology, 08860 Castelldefels (Barcelona), Spain}
\affiliation{ICREA, Passeig Lluis Companys 23, 08010 Barcelona, Spain}

\author{C. Psaroudaki}
\affiliation{Laboratoire de Physique de l’\'{E}cole Normale Sup\'{e}rieure, ENS, Universit\'{e} PSL, CNRS, Sorbonne Universit\'{e}, Universit\'{e} de Paris, F-75005 Paris, France}
\author{K. Roszak}
\affiliation{FZU - Institute of Physics of the Czech Academy of Sciences, 182 00 Prague, Czech Republic}

\date{\today}
\begin{abstract}
Extracting information from quantum many-body systems remains a key challenge in quantum technologies due to experimental limitations. In this work, we employ a single spin qubit to probe a strongly interacting system, creating an environment conducive to qubit decoherence. By focusing on the XXZ spin chain, we observe diverse dynamics in the qubit evolution, reflecting different parameters of the chain. This demonstrates that a spin qubit can probe both quantitative properties of the spin chain and qualitative characteristics, such as the bipartite entanglement entropy, phase transitions, and perturbation propagation velocity within the system. This approach reveals the power of small quantum systems to probe the properties of large, strongly correlated quantum systems. 

\end{abstract}
\maketitle
%------------------------------------------------------------------------------------ 

{\it Introduction ---} Characterizing quantum phases in strongly correlated systems has been a central focus of condensed matter theory. In general, quantum phase classification is based on a plethora of quantities: from ground state energy gap and its entanglement entropy to higher-order correlation functions, all of which depend on Hamiltonian parameters. However, accessing these quantities experimentally is often challenging, creating a strong demand for methods that characterize strongly correlated quantum phases through single-particle measurements.

Single-qubit decoherence dynamics offer a pathway for extracting information about a given quantum system. The qubit coupled to the quantum system in a controlled way serves as a probe and the decoherence dynamics can provide information about the properties of the system of interest.
Decoherence dynamics is used for quantum sensing and noise spectroscopy to explore the fundamental properties of quantum systems and to measure physical quantities with high sensitivity \cite{clerk10,young12,szankowski17,degen17,PhysRevLett.112.223601}.
Notably, simple two-level systems can act as efficient spectrum analyzers of quantum noise, typically determined by its impact on measurable quantities, such as the decoherence rate of a qubit \cite{PhysRevLett.92.117905,PhysRevA.86.012314}. Single-qubit sensing techniques offer insights into specific features of the noise spectra \cite{yan13,hall16}, which can be used to infer properties of the environment \cite{degen17,jerger23} and intrinsic aspects of interactions, such as entanglement generation \cite{roszak19a,rzepkowski21,strzalka21,zhan21}. 

\begin{figure}[t!]
	\includegraphics[scale=0.7]{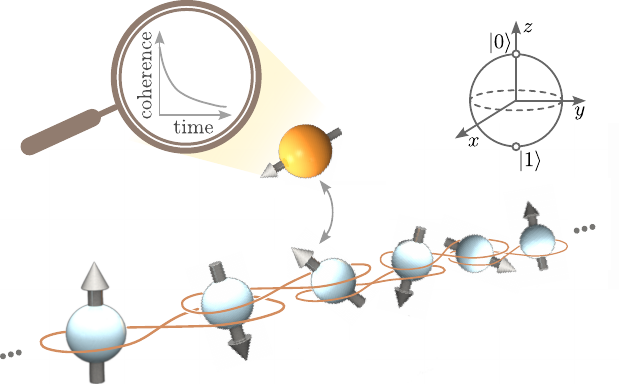}	\caption{The spin-qubit probe coupled to a strongly correlated quantum spin chain. The chain acts as an environment for the qubit, which facilitates decoherence, while
		 the qubit introduces a perturbation to the chain propagating in the system. The propagation depends on system properties and consequently determines qubit dynamics. Thus the information about the chain becomes encoded in the coherence dynamics, which can be easily measured and analyzed.
 \label{fig1}}
\end{figure}

So far, single-qubit sensing has been demonstrated across various platforms, including trapped atoms \cite{almog11} and ions \cite{biercuk09}, superconducting qubits \cite{bylander11}, molecular spins \cite{fu21}, and quantum dots \cite{chan18}. Among these, single-spin magnetometers based on solid-state electronic spin defects, such as nitrogen-vacancy (NV) centers in diamond \cite{101146,Casola2018} or boron vacancies in hexagonal boron nitride (hBN) \cite{Healey2023}, are sensitive quantum sensors for detecting weak magnetic signals. With efficient initialization, readout capabilities, and relatively long coherence times, quantum impurities serve as effective tools for noninvasively measuring collective excitations in magnetic insulators \cite{PhysRevLett.121.187204, vanderSar2015}. They can probe magnetic field noise \cite{Park2022}, detect nonclassical magnetic ground states \cite{PhysRevLett.131.143602}, and diagnose various phases of magnetic insulators \cite{PhysRevLett.131.070801}, including spin liquids with fractionalized excitations \cite{PhysRevB.99.104425}.

In noise spectroscopy, it is typically assumed that the interactions within the probed system are weak \cite{roy15, cetina16, fukami21}.
Under these conditions, the Born-Markov approximation, or
the equivalent Fermi’s golden rule, is often applied \cite{flebus18, chatterjee19},  
leading to an exponential decoherence decay characterized uniquely by relaxation and dephasing times. This strong assumption is known to fail commonly already in the case of non-interacting systems \cite{nakatani10,breuer16,devega17,kolovsky20,lonigro22}, while the limits of its applicability are not well understood with some notable exceptions, such as the spin-boson model \cite{reina02,tuziemski19,PhysRevB.107.085428}.

In this work, we use a spin-qubit to probe a strongly correlated many-body quantum system, modeled as a one-dimensional spin-$1/2$ chain, see Fig.\ref{fig1}. 
From the qubit perspective, the spin-chain constitutes an environment and causes decoherence. For the chain,
the coupling to the qubit introduces a perturbation, the propagation of which is determined by system properties. This in turn determines the spin-qubit decoherence dynamics and facilitates the transfer of information about the chain into the evolution of the qubit. We show that the qubit decoherence can serve as a probe of quantum phase boundaries, offering valuable insights into the velocity of information propagation.

We consider the one-dimensional Heisenberg XXZ model \cite{PhysRev.147.303,PhysRev.150.321,PhysRev.135.A640,PhysRevLett.59.259,giamarchi04,franchini17,marzolino17}.
The XXZ Hamiltonian is an archetypical many-body quantum system, exactly solvable with Bethe ansatz, with well-established ground state properties \cite{giamarchi04,franchini17}. These properties are governed by the anisotropy parameter $\Delta$, which tunes the relative strength of interactions between neighboring spins depending on their alignment in the $z$ versus the $x,y$ direction. In the thermodynamic limit of an infinite system, for $-1 < \Delta \le 1$ the ground state of the XXZ system is the gapless XY-phase with quasi-long-range order in the XY-plane and power-law decay of spin correlations \cite{BraiorrOrrs2016}. For $\Delta\le-1$ the ground state is in the gapped ferromagnetic phase, while for $\Delta>2$ in the gapped antiferromagnetic phase.
 At the isotropic point $\Delta=1$, the model is realized in several one-dimensional AFM meterials including Sr$_2$CuO$_3$, SrCuO$_2$, CaCu2O$_3$, and YbMgGaO$_4$ crystals \cite{PhysRevLett.76.3212,Paddison2017,HESS20191}. For $\Delta=1/2$, is serves as a low energy effective model for the magnetic-field induced gapless phase of $S=1$ NiCl$_2$ -SC(NH$_2$ )$_2$ \cite{PhysRevB.89.224418}. 
 The XXZ spin chain with arbitrary $J$ and $\Delta$ can be realized in an arrays of Rydberg atoms \cite{Toskovic2016,PhysRevX.8.011032,PhysRevX.14.011025}, as well as with trapped polar molecules \cite{Yue2022}.

\begin{figure*}[t]
    \centering
       \includegraphics[width=\linewidth]{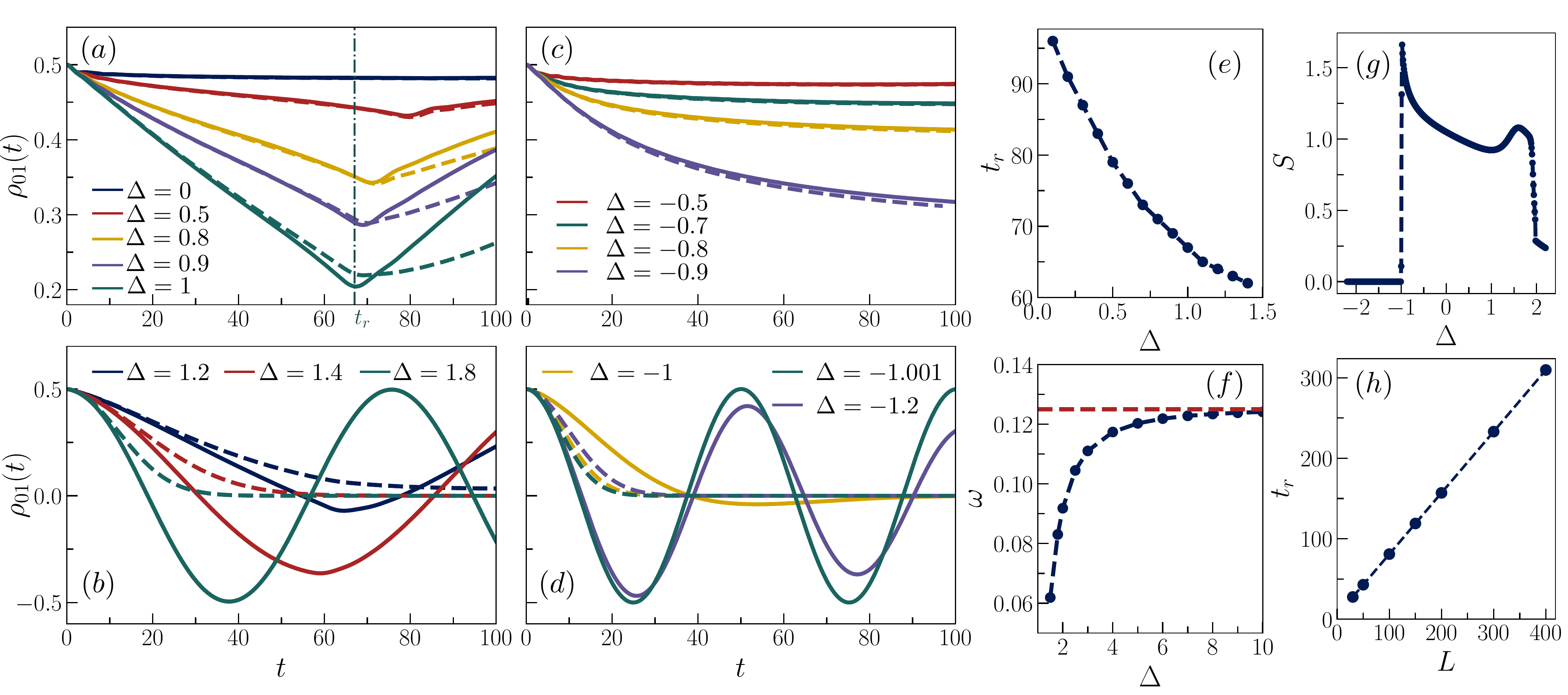}
    \caption{
    Left: Evolution of qubit coherence for different values of the anisotropy parameter
    (TDVP - solid lines, TCL - dashed lines):
    (a) positive $|\Delta|\le 1$; (b) positive $|\Delta|> 1$;
    (c) negative $|\Delta|\le 1$; (d) negative $|\Delta|> 1$.
    Right: Dependence of (e) the qubit recoherence time $t_r$, (f) oscillation frequency $\omega$, (g) and entanglement entropy $S$ on anisotropy parameter $\Delta$. Dependence of qubit recoherence time $t_r$ on $L$ (h).}
    \label{Fig:Panel}
\end{figure*}

{\it Model---}
The system consists of a spin-qubit coupled to the spin-$1/2$ chain with $L$ spins via a local interaction. The total Hamiltonian reads $H = H_Q + H_S + H_{SQ}$. $H_\mathrm{Q}=h_z\,\sigma^z_\mathrm{Q}/2$ is the free qubit Hamiltonian, where $\sigma^z_\mathrm{Q}$ is the Pauli matrix and $h_z$ is the qubit energy level splitting.
The Hamiltonian $H_S$ for spin-$1/2$ chain is given by the XXZ model \cite{giamarchi04},
%\begin{equation}
%\label{xxz}
	$H_\mathrm{S}=J\sum_{i=1}^{L}\left[\frac{1}{2}\left(S^{+}_{i} S^{-}_{i+1}+\mathrm{H.c.}\right)+\Delta S^z_{i} S^z_{i+1}\right]$,
%\end{equation}
where  $J>0$, and $\Delta$ is the anisotropic parameter.  $S^{+,-,z}_{i}$ are standard spin-$1/2$ operators acting on the $i$-th spin; we assume open boundary conditions.
Finally, the qubit is coupled only to the middle spin of the chain via $H_\mathrm{SQ}=g\,S^z_{L/2} \sigma^z_\mathrm{Q}/2$, where $g$ is the coupling strength. The commutation of $H_Q$ and $H_{SQ}$ implies that only the qubit coherence is disturbed by the chain, so the qubit evolution is restricted to pure decoherence \cite{roszak15}.

The initial state of the total system is a product state of the XXZ chain ground state and the equal superposition state of the qubit, $\frac{1}{\sqrt{2}}(|0\rangle + |1\rangle)$. We numerically evolve the initial state of the composite system and analyze the time evolution of the qubit coherence, $\rho_{01}(t) = \langle 0|\rho_Q(t)|1\rangle$, where $\rho_Q(t)$ is reduced density matrix of the qubit.
The chain ground state is obtained via the density matrix renormalization group (DMRG) technique \cite{PhysRevLett.69.2863,PhysRevB.48.10345,RevModPhys.77.259,SCHOLLWOCK201196,ORUS2014117}. The full-time evolution of the qubit is obtained with a time-dependent variational principle scheme (TDVP) \cite{Kramer_2008, PhysRevLett.107.070601, PhysRevB.94.165116,PhysRevB.102.094315}. 
We also find an approximate qubit evolution using the time-convolutionless projection operator (TCL) method
\cite{breuer02} which allows us to reliably quantify the observed changes in the qubit evolution.

% \begin{figure}[t!]
%     \centering
    
%      \includegraphics[width=\linewidth]{fig_2.png}
%     \caption{Ground state energy gap (black solid line) and entanglement entropy (red marks) for the XXZ ground state as a function of the anisotropy parameter for finite spin-$1/2$ chain with $L = 100$ and open boundary conditions. For $\Delta<-1$ the XXZ ground state is a gapless with a spin-$1/2$ product structure. With increasing anisotropy paramter, at sharp transition, $\Delta=-1$, system enters the finite gap phase and finite bipartite entanglement entropy. For $\Delta\gtrsim1.8$ system has smoothly vanishing energy gap, and smoothly decaying entanglement.}
%     \label{fig2}
% \end{figure}
 
{\it Results ---}
The solid lines in Fig.~\ref{Fig:Panel} show the evolution of the probe
qubit coherence $\rho_{01}(t)$ obtained via TDVP. Here we set $J=1$, and various $\Delta$ anisotropy values corresponding to different quantum phases, while the coupling constant is fixed at $g=1/4$ and the number of spins to $L=100$. The energy splitting of the probe qubit, $h_z$, is set to zero, keeping the qubit coherence $\rho_{01}(t)$ remaining real throughout the evolution. Note that taking $h_z\neq 0$ only yields unitary qubit oscillations.

Fig.~\ref{Fig:Panel} (a) and (c) show the $|\Delta|\le 1$ results for positive and negative $\Delta$, respectively.
In this regime, the initial fast decoherence is followed by a slower dephasing, the magnitude of increases with $\vert \Delta \vert$, indicating that stronger interactions within the chain result in more pronounced decoherence of the probe qubit. With increasing $\vert \Delta \vert$, the longitudinal correlations along the $z$-axis become more pronounced relative to the XX model at $\Delta=0$, where only the in-plane interactions are relevant. There is however a difference in character between the negative and positive $\Delta$ evolutions, as we discuss below.  

Fig.~\ref{Fig:Panel} (a) suggests that the spin-qubit coherence can serve as a probe for the speed of information propagation in the XXZ chain for positive $\Delta\le 1$. This is evident through the qubit's recoherence at time $t_r$ (highlighted by the vertical dashed line for $\Delta =1$). The finite $t_r$ corresponds to the reflection of the qubit-induced perturbation from the ends of the XXZ chain. Recoherence time decreases with $\Delta$, see Fig.~\ref{Fig:Panel} (e), indicating that qubit coherence dynamics contain information about the velocity of perturbation propagation \cite{Lieb1972,PrmontSchwarz2010,wysocki2024}. For negative $\Delta\ge -1$ there is no recoherence due to boundary effects observable for the considered timescales [see Fig.~\ref{Fig:Panel} (c)]. This suggests that information propagation velocity is significantly lower compared to the case of positive $\Delta\le 1$. Indeed, at zero temperature and in the gapless phase of the XXZ model, the perturbation caused by the spin qubit is mediated by spinons, the elementary fractional excitations of the chain. The spinon travels with a velocity $u_s=J \pi \sqrt{1-\Delta^2}/2 \cos^{-1}\Delta$ \cite{Takahashi_1999}, and is reabsorbed by the spin qubit after a time $t_r$, which decreases with increasing $\Delta$, as shown in Fig.~\ref{Fig:Panel} (e). The increase in the $z$-component correlation functions and the rise in spinon velocity with $\Delta$ reflect the same underlying mechanism: the system becomes more ordered along the $z$-direction, with interactions becoming more robust, a behavior captured in the time evolution of the spin qubit. Finally, as expected, the recoherence time depends on the system size and is proportional to $L$, depicted in Fig.~\ref{Fig:Panel} (h). 

Fig.~\ref{Fig:Panel} (b) shows the qubit evolution for the chain above the antiferromagnetic isotropic point,  $\Delta>1$. We observe a gradual qualitative change in the evolution, which culminates when the decaying character characteristic of $\Delta\le 1$ is replaced by periodic evolution for $\Delta>1.8$. In this regime, the system enters its gapped phase with a quasi-degenerate ground state with $S^z=0$ \cite{10.1063/1.1705048}, but finite sublattice magnetization and finite entanglement (see Fig.~\ref{Fig:Panel} (g)). These states are remnants of the classical AFM ordered states and turn into the degenerate Ne\'{e}l states with vanishing entanglement in the Ising limit $\Delta \rightarrow \infty$. For the finite chain of $L=100$ sites considered here, the classical limit is reached at $\Delta=1.8$. Above this point, the oscillation frequency $\omega$ of the periodic spin-qubit evolution further increases with $\Delta$ and saturates for large $\Delta \gg 1$ at $\omega = g/2 = 0.125$ [see  Fig.~\ref{Fig:Panel} (f)]. The detection of the change at $\Delta\approx 1$ and analysis of the intermediate $1<\Delta<1.8$ and $\Delta \gg 1$ regime, requires input from open quantum system methodology and is discussed below. 

The behavior of the qubit probe for negative $\Delta<-1$, when the system is gapped and unentangled, is presented in Fig.~\ref{Fig:Panel} (d). For any system size, the system is in the ferromagnetic Ising phase with a saturated ground state with all spins aligned in $\pm z$-direction with total magnetization $S^z=\pm L/2$. Here the transition to oscillatory qubit evolution occurs sharply at $\Delta=-1$, with no transition
phase (see SM) and the frequency of oscillations is independent of $\Delta$,  $\omega=0.05$. 
The transition from decaying to oscillatory qubit evolution for positive and negative $\Delta$ values,
reflects the difference in the transition between the entangled and unentangled ground state. 

For any $\Delta$, the decay of initial qubit coherence is not exponential, implying that non-Markovian effects are
always non-negligible in the studied scenario. These results offer compelling evidence that a controllable single qubit can act as a highly sensitive probe, capable of detecting quantum phase transitions associated with the sudden vanishing of ground state entanglement near the phase boundary, as well as providing information about the velocity of perturbation propagation in the XXZ chain. For completeness, we present the decoherence dynamics for larger coupling values in the Supplementary Materials (SM). As anticipated, $\rho_{01}(t)$ decays more rapidly as the coupling strength $g$ increases.

{\it TCL analysis ---}
To get insight into the physical processes governing the spin-qubit coherence dynamics after coupling to a strongly correlated quantum system, we employ the semi-analytical framework of second-order TCL projection operator expansion \cite{breuer02,roszak09,barnes12}. 
The TCL analytical predictions offer a clear interpretation of the processes influencing the qubit evolution, enabling distinctions between single and many-particle processes, transitions and scattering events, and non-Markovian and Markovian effects. TCL provides a way to approximate the reduced density matrix using a perturbation expansion with respect to the strength of the interaction Hamiltonian $g$. Since the qubit-spin-chain initial state is factorized and the odd moments of the interaction Hamiltonian vanish with respect to the initial ground state of the XXZ model, we obtain a
master equation equivalent to the Redfield equation \cite{redfield57}, which is time-local, but not Markovian
(the derivation can be found in SM). The solution of this equation yields the time evolution of the probe qubit coherence
in the interaction picture and is given by
\begin{eqnarray} \label{rho2}
\rho_{01}^{\mathrm{2nd}}(t) & =&\rho_{01}(0)\exp\left(-\frac{g^2}{2}\int\limits_0^{t}\mathrm{d}\tau\,A(\tau)\right),\\
\nonumber
	A(\tau) & =&\int_{-\tau}^{\tau}\mathrm{d}t\, \langle S^z_{L/2}(t)S^z_{L/2}(0)\rangle,
\end{eqnarray}
where $S^z_{L/2}(t)=e^{iH_\mathrm{S}t}S^z_{L/2}e^{-iH_\mathrm{S}t}$
and the expectation value is taken with respect to the initial (ground) state of the spin-chain. The coherence decay of Eq.~(\ref{rho2}) depends on the two-time correlation function of the middle spin of the XXZ chain, which is obtained numerically using TDVP. The function describes decoherence which is dictated by single-particle processes on the spin coupled to the probe qubit. The Markov limit, where the memoryless dynamics of the spin-qubit does not depend on the history of its interaction with the environment, is recovered in the $\tau \rightarrow \infty$ limit. This approximate solution is incapable of accounting for higher-order processes. Note that the interaction picture results are equivalent to $h_z=0$.

The decoherence evolution obtained from Eq.~(\ref{rho2}) is plotted in Fig.\ref{Fig:Panel} (a-d) by dashed lines. For $|\Delta|\le 1$, panels (a) and (c), the TDVP and TCL curves are in almost perfect agreement, both qualitatively and quantitatively. Even above the point where the finite-chain-length effects become significant,
the agreement is remarkably good. Above the antiferromagnetic isotropic point of the chain $\Delta>1$ [Fig.~\ref{Fig:Panel} (b)],
a discrepancy between the two types of results begins to emerge, indicating that higher-order processes have become significant. With growing $\Delta$ the discrepancy grows and nearing $\Delta = 1.8$, the oscillatory character of the decoherence becomes dominant. The same holds for $\Delta<-1$, where TCL fails to reproduce the oscillations of the qubit evolution. To understand the observed changes in probe qubit behavior and validate our results for an infinite XXZ chain, we now analytically examine the two limiting cases of the XXZ Hamiltonian with $\Delta = 0$ and $\Delta \rightarrow \pm \infty$.

{\it Analytical TCL results for $\Delta=0$ ---}
While deriving the complete solution for the qubit evolution with $\Delta = 0$ is analytically challenging \cite{giamarchi04,popovic23}, obtaining the second-order TCL approximate solution of Eq.~\eqref{rho2} is straightforward via the Jordan-Wigner transformation in Fourier space \cite{giamarchi04}. The Hamiltonian reads $H_{\mathrm{S}}^{\mathrm{0}}=(J/4)\sum_k\cos(k)c_k^{\dagger}c_k$, where $k=2\pi n/L$ with $n=1,2,\dots,L$, and $2S^{-}_{i}=(1/\sqrt{L})\sum_ke^{ikr_i}c_k$.
Since the ground state is the filled Fermi sea, it can be written as $|g\rangle = \prod_{k<k_F}c_k^{\dagger}|0\rangle$,
where $k_F$ is the wave vector corresponding to the Fermi energy (which is zero) and $|0\rangle$ denotes a state
with no excitations. Hence we get $\langle S^z_{M}(t)S^z_{M}(0)\rangle=(4/L^2)\sum_{k\ge k_F,k'<k_F}\exp\left[
i(\varepsilon(k')-\varepsilon(k))t\right]$, with $\varepsilon(k)=(J/4)\cos(k)$. Consequently, the qubit evolution is given by
\begin{equation}
\label{rho20}
\frac{\rho_{01}^{\mathrm{2nd}}(t)}{\rho_{01}(0)}=\exp\left(\frac{4g^2}{L^2}\sum_{k\ge k_F,k'<k_F}
\frac{\cos\left((\varepsilon(k')-\varepsilon(k))t\right)-1}{(\varepsilon(k')-\varepsilon(k))^2}
\right).
\end{equation}

Several insights can be drawn from Eq.~(\ref{rho20}). Firstly, qubit decoherence in this regime arises from numerous single-fermion transitions, involving excitations with varying momenta. Thus, from the qubit's perspective, the environment behaves as a large system in terms of the number of fermionic degrees of freedom. Secondly, in the Markov limit, $\tau\rightarrow\infty$ in Eq.~(\ref{rho2}), no decoherence occurs and $\rho_{01}^{\mathrm{2nd}}(t)=\rho_{01}(0)$. Finally, this formula can be readily extended to an infinite chain, showing no qualitative changes in the nature of the decoherence. The correspondence of the TDVP and TCL curves confirms that the chain acts as a many-body environment for the qubit, while the non-Markovian nature of the decoherence is indisputable. Summarizing, for $|\Delta| \le 1$, the probe qubit detects a many-body environment,
which can be successfully described with only  single particle qubit-chain processes, as long as non-Markovian effects are taken into account.  

{\it Analytical results for $\Delta\to\pm\infty$ ---} 
To gain insight into the qubit evolution in this regime we examine the Ising Hamiltonian for the chain,
$
H_\mathrm{S}^\mathrm{I}=J\sum_{i=1}^L\Delta S^z_{i} S^z_{i+1}.
$
In this case, the whole system Hamiltonian $H=H_\mathrm{Q}+H_\mathrm{S}^\mathrm{I}+H_\mathrm{QS}$, is diagonal in the $\sigma^z$ basis.
The ground state of the Ising chain is degenerate and encompasses the two antiferromagnetic configurations, 
$|\!\!\!\uparrow\downarrow\rangle\equiv|\!\!\!\uparrow\downarrow\uparrow\downarrow\!\!\dots\rangle$,
$|\!\!\!\downarrow\uparrow\rangle\equiv|\!\!\!\downarrow\uparrow\downarrow\uparrow\!\!\dots\rangle$. Any superposition of these states is also an eigenstate of the Ising Hamiltonian, furthermomre these states are also eigenstates of 
the full Hamiltonian, and cannot
lead to qubit decoherence. Contrarily, any superposition of these states is not an eigenstate of the interaction Hamiltonian due to a difference in symmetry between the Hamiltonian components, and always leads to qubit dephasing. 
The evolution of the qubit-chain system for the initial chain state, $|\psi_{\rm ini}\rangle=\alpha|\!\uparrow\downarrow\rangle+\beta|\!\downarrow\uparrow\rangle$,
simply reads
% \begin{eqnarray}
% \nonumber
% |\psi_{\mathrm{QS}}(t)\rangle &=&\frac{1}{\sqrt{2}}\left[e^{-i\frac{h_z}{2}t}|0\rangle\otimes
% \left(\alpha e^{-i\frac{g}{2}t}|\!\uparrow\downarrow\rangle+\beta e^{i\frac{g}{2}t}|\!\downarrow\uparrow\rangle
% \right)\right.\\
% \label{psi}
% &&+e^{i\frac{h_z}{2}t}|1\rangle\otimes\left.
% \left(\alpha e^{i\frac{g}{2}t}|\!\uparrow\downarrow\rangle+\beta e^{-i\frac{g}{2}t}|\!\downarrow\uparrow\rangle
% \right)
% \right].
% \end{eqnarray}
$|\psi_{\mathrm{QS}}(t)\rangle =\frac{1}{\sqrt{2}}\left[e^{-i\frac{h_z}{2}t}|0\rangle\otimes
\left(\alpha e^{-i\frac{g}{2}t}|\!\uparrow\downarrow\rangle+\beta e^{i\frac{g}{2}t}|\!\downarrow\uparrow\rangle
\right)\right.
\label{psi}
+e^{i\frac{h_z}{2}t}|1\rangle\otimes\left.
\left(\alpha e^{i\frac{g}{2}t}|\!\uparrow\downarrow\rangle+\beta e^{-i\frac{g}{2}t}|\!\downarrow\uparrow\rangle
\right)
\right]$.
The $\Delta\to \infty$ limit reduces the chain to an effective two-level system, and the spin-qubit coherence evolution in the interaction picture reads
%\begin{equation}
%\label{rhoi}
%\rho_{01}(t)=\rho_{01}(0)\left(|\alpha|^2e^{-igt}+|\beta|^2e^{igt}\right),
%\end{equation}
$\rho_{01}(t)=\rho_{01}(0)\left(|\alpha|^2e^{-ig/2t}+|\beta|^2e^{ig/2t}\right)$
yielding a function that is purely real only if $|\alpha|^2=|\beta|^2=1/2$.
All TDVP results for the XXZ chain are real at all times, which justifies using the equal superposition initial state of the Ising chain. 
The simplified probe qubit evolution is purely cosinusoidal, depending only on the qubit-chain coupling strength $g$, consistent with the TDVP result of the oscillation frequency converging to $\omega=g/2$ in the $\Delta \gg 1$ limit, as shown in Fig.~\ref{Fig:Panel} (f). The Ising chain with negative $\Delta$ yields analogous results. Here the qubit evolution is also oscillatory, but the ground states of the chain are ferromagnetic,
	$|\!\!\!\uparrow\uparrow\rangle\equiv|\!\!\!\uparrow\uparrow\uparrow\uparrow\!\!\dots\rangle$, and
	$|\!\!\!\downarrow\downarrow\rangle\equiv|\!\!\!\downarrow\downarrow\downarrow\downarrow\!\!\dots\rangle$. Hence, for $\Delta\rightarrow\pm\infty$, the spin chain effectively provides a two-level environment for the qubit, leading to cosinusoidal evolution of the spin-qubit coherence as well.

{\it Conclusions ---} We have shown that a single spin-qubit can serve as a probe for the study of a strongly interacting XXZ spin-$1/2$ chain. Qubit coherence exhibits high sensitivity to the chain's Hamiltonian parameters, encoding information about the global properties of the spin chain. We demonstrate that a controllable single qubit can detect quantum phase transitions associated with the sudden vanishing of the ground state entanglement near the phase boundary, and provide information about the velocity of perturbation propagation along the chain.

Depending on the parameters, the qubit displays non-Markovian decoherence curves characteristic of a many-body environment which corresponds to the Born approximation, or oscillatory behavior in which the chain effectively acts as a two-level system. An intermediate behavior is also observed, where the Born approximation no longer holds, but the spin chain still fulfills the role of a many-body environment. Transitions between these decoherence regimes correspond unambiguously to the properties of the ground state bipartite entanglement entropy. 
Detecting phase transition requires measuring only a single qubit, irrespective of the size of the spin chain. This ensures that the method scales efficiently as the system size increases.
These observations highlight the potential of small quantum systems for probing properties of large, strongly correlated quantum systems.

{\it Acknowledgments ---}
K.R. is grateful to dr.~Jacek Herbrych for early discussions. S.D. thanks dr.~Hari Kumar Yadalam for helpful discussions. We thank Weronika Golletz for preparing Fig.~1 in the manuscript.

M.P. and S.D. had equal contributions to the project. 

S.D. acknowledges support from the European Union and the Czech Ministry of Education, Youth and Sports (MEYS)(Project: MSCA Fellowship CZ FZU III – CZ.02.01.01/00/22010/0008598) and the computational resources provided by the e-INFRA CZ project (ID:90254), also supported by MEYS.
C.P. is an \'{E}cole Normale Sup\'{e}rieure (ENS)-Mitsubishi Heavy Industries (MHI) Chair of Quantum Information supported by MHI. 
K.R. acknowledges support from COST Action
SUPERQUMAP, supported by COST
(European Cooperation in Science and Technology).

ICFO group acknowledges support from:
European Research Council AdG NOQIA; MCIN/AEI (PGC2018-0910.13039/501100011033, CEX2019-000910-S/10.13039/501100011033, Plan National FIDEUA PID2019-106901GB-I00, Plan National STAMEENA PID2022-139099NB, I00, project funded by MCIN/AEI/10.13039/501100011033 and by the “European Union NextGenerationEU/PRTR" (PRTR-C17.I1), FPI); 
QUANTERA MAQS PCI2019-111828-2; QUANTERA DYNAMITE PCI2022-132919, QuantERA II Programme co-funded by European Union’s Horizon 2020 program under Grant Agreement No 101017733; Ministry for Digital Transformation and of Civil Service of the Spanish Government through the QUANTUM ENIA project call - Quantum Spain project, and by the European Union through the Recovery, Transformation and Resilience Plan - NextGenerationEU within the framework of the Digital Spain 2026 Agenda; Fundació Cellex; Fundació Mir-Puig; Generalitat de Catalunya (European Social Fund FEDER and CERCA program, AGAUR Grant No. 2021 SGR 01452, QuantumCAT \ U16-011424, co-funded by ERDF Operational Program of Catalonia 2014-2020); Barcelona Supercomputing Center MareNostrum (FI-2023-3-0024);  Funded by the European Union. 

Views and opinions expressed are however those of the author(s) only and do not necessarily reflect those of the European Union, European Commission, European Climate, Infrastructure and Environment Executive Agency (CINEA), or any other granting authority.  Neither the European Union nor any granting authority can be held responsible for them (HORIZON-CL4-2022-QUANTUM-02-SGA  PASQuanS2.1, 101113690, EU Horizon 2020 FET-OPEN OPTOlogic, Grant No 899794),  EU Horizon Europe Program (This project has received funding from the European Union’s Horizon Europe research and innovation program under grant agreement No 101080086 NeQSTGrant Agreement 101080086 — NeQST);  ICFO Internal “QuantumGaudi” project;  European Union’s Horizon 2020 program under the Marie Sklodowska-Curie grant agreement No 847648; “La Caixa” Junior Leaders fellowships, La Caixa” Foundation (ID 100010434): CF/BQ/PR23/11980043.

%================================================================================
\bibliography{bibliography} 

~\\
\newpage
\section{Supplemental Material for:\\H\lowercase{arnessing spin-qubit decoherence to probe strongly-interacting quantum systems}}

\subsection{Time-convolutionless projection operator method}
Projection operator techniques are important tools for studying the behavior of complex open quantum systems~\cite{breuer02}. The generalized quantum 
master equation is a powerful method for describing the dynamics of quantum systems that are not in equilibrium. One well-known approach within these 
techniques is the Nakajima–Zwanzig (NZ) time-convolution projection operator method~\cite{Nakajima,Zwanzig}. This method simplifies the challenging 
task of solving the many-body quantum Liouville-Von Neumann equation by focusing on a super-operator called the `memory kernel.' The memory kernel 
contains all the information needed to understand the non-Markovian dynamics of the system. In simpler terms, it allows us to study how the system's 
past influences its present and future behavior.

An alternative to the NZ generalized quantum master equation is the time-convolutionless (TCL) quantum master equation~\cite{Shibata1977, Hashitsume1977,
Chaturvedi1979,Shibata1980}. The TCL approach has the advantage of providing a time-local equation of motion for the relevant parts of the system. This 
means that the system's evolution does not depend on its history, making the equation simpler to solve. Instead of dealing with the complex NZ memory 
kernel, the TCL method uses a time-local super-operator called the TCL generator or kernel. However, directly calculating the TCL generator is  
challenging because it involves finding an inverse super-operator in the full Hilbert space (the complete mathematical space describing the system). 
Therefore, the TCL generator is typically obtained using a perturbative approach, which simplifies calculations by considering small system-bath interactions, and this method has been successfully applied to many physical systems~\cite{Breuer1999,Breuer2004,Breuer2006,Krovi2007}.

In this paper we apply the TCL projection operator technique to the model of a single spin qubit (Q) interacting with a spin chain (which
works as the environment for the qubit in the context of the TCL calculations) of N spins as shown in Fig.~1 of the main text. The model Hamiltonian corresponding to this composite system (qubit+chain) is written as
\begin{equation}
	H = H_{Q}+H_{S}+H_{SQ}.
\end{equation}
Here $H_{0}=H_{Q}+H_{S}$ is the unpertured Hamiltonian, and $H_{SQ}$ is the interaction between qubit and chain. 
In the interaction picture with respect to $H_{0}$, we obtain the time-dependent Hamiltonian,
\begin{equation}
\label{pict}
	H_{SQ}(t) = e^{iH_{0}t} H_{SQ} e^{-iH_{0}t} 
		 = g   S_{M}^z(t)\sigma_{Q}^z/2,
\end{equation}
where the time-dependence is only present in the chain operators due to the commutation of $H_{Q}$ and $H_{SQ}$, and $S_{M}^z(t)=e^{iH_{S}t}S_{M}^ze^{-iH_{S}t} $.
Note that $S_{M}^z$ is the spin operator corresponding to the middle spin on the chain, the one which is interacting with the qubit.

\begin{figure}
\includegraphics[width=\linewidth]{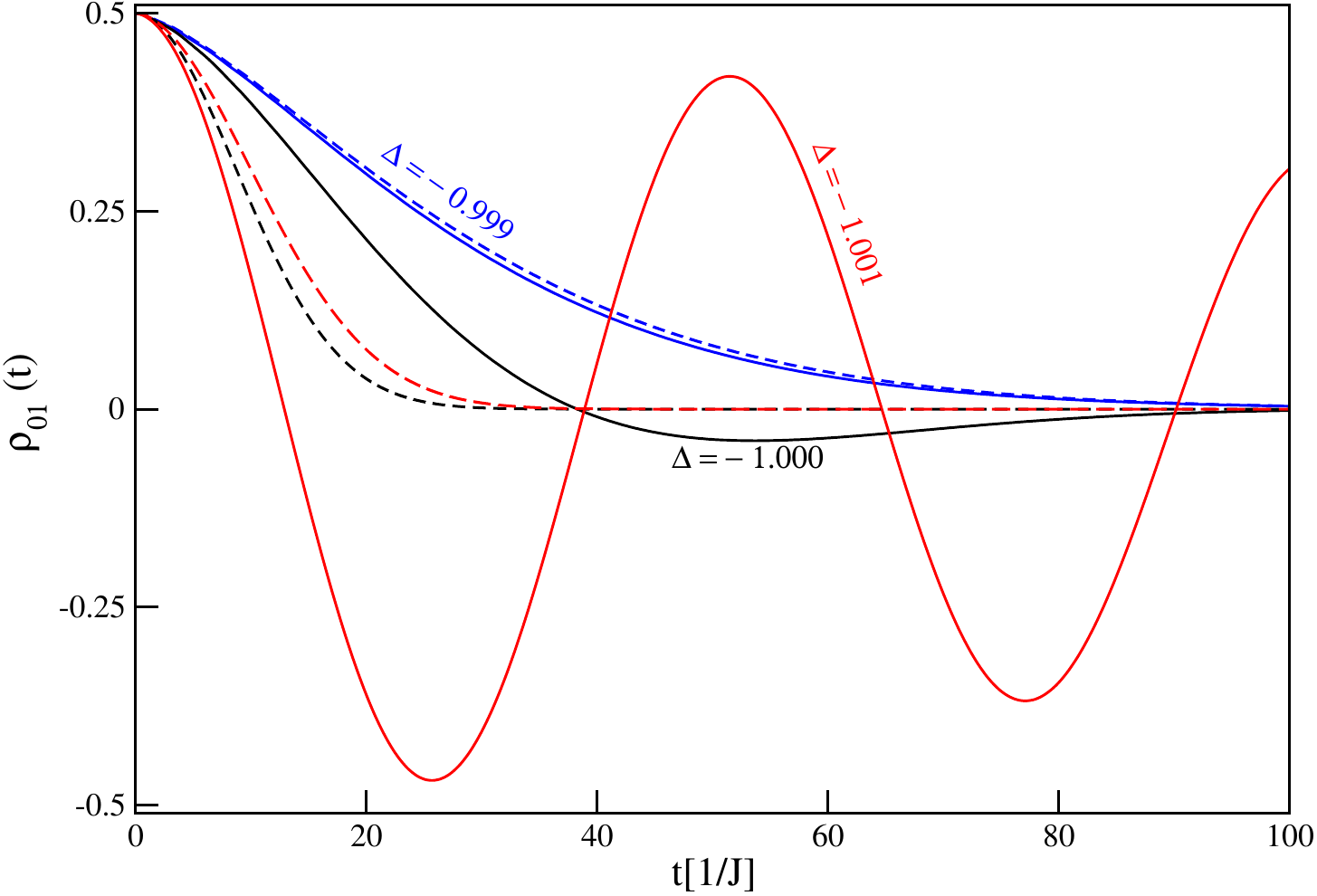}
\caption{(a) Qubit decoherence for different values of $\Delta\approx -1$ (solid lines - TDVP, dashed lines - TCL).
}
\label{fig_SM1}
\end{figure}

Starting from the Liouville-von Neumann equation for the evolution of the qubit+chain density matrix, $\rho_{SQ}$, it is possible to derive the exact form 
of the TCL equation \cite{breuer02}. This is done using the superoperators $\mathcal{P}$ and $\mathcal{Q}$, which project $\rho_{SQ}$ into relevant and irrelevant subspaces
as follows
\begin{eqnarray}
\mathcal{P}\rho_{SQ}(t)&=&\rho_{Q}(t)\otimes\rho_S(0),\\
\mathcal{Q}&=&I-\mathcal{P},
\end{eqnarray}
so that $\mathrm{Tr}_S\mathcal{P}\rho_{SQ}(t)=\rho_{Q}(t)$, which is the quantity of interest. $I$ denotes the unit operation. 
The TCL equation, which is time local in time, contrarily to the NZ equation, is of the form
\begin{equation}
\label{tcleq}
\frac{d}{dt} \mathcal{P}\rho_{SQ}(t) = \mathcal{K}(t)\mathcal{P}\rho_{SQ}(t)+\mathcal{I}(t)\mathcal{Q}\rho_{SQ}(0).
\end{equation}
$\mathcal{K}(t)$ is the so called the TCL generator, while $\mathcal{I}(t)$ is the inhomogeneity. The inhomogeneity acts on the irrelevant part 
of the initial qubit-chain density matrix, so it describes the effect of correlations between the two subsystems initially present on the evolution
of the system. Since for product initial
conditions $\mathcal{Q}\rho_{SQ}(0)=0$, we do not need to take into account the last term in eq.~(\ref{tcleq}).

Expansion of the TCL generator, $\mathcal{K}(t)$, is done with the help of the equations for the relevant and irrelevant parts of the density matrix,
as well as propagators that describe the evolution both forwards and backward in time \cite{breuer02}.
Truncating at second order, we get
$\mathcal{K}(t) = \mathcal{K}^{(1)}(t)+\mathcal{K}^{(2)}(t)$, where the index denotes the order. The two terms are given by
\begin{eqnarray}
\mathcal{K}^{(1)}(t) &=&\mathcal{P}\mathcal{L}(t)\mathcal{P},\\
\mathcal{K}^{(2)}(t) &=& \int_0^t ds \left[\mathcal{P}\mathcal{L}(t)\mathcal{L}(s)\mathcal{P}-\mathcal{P}\mathcal{L}(t)\mathcal{P}\mathcal{L}(s)\mathcal{P} \right]
\end{eqnarray}
Here $\mathcal{L}(t)$ is the Liouville superoperator and it is defined by 
\begin{equation}
\label{liu}
\mathcal{L}(t) \rho(s) = -i\left[ H_{SQ}(t), \rho_{SQ}(s) \right].
\end{equation}

For many systems, where the environment (chain in our case) is initially at equilibrium we have 
\begin{equation}
\label{plp}
\mathcal{P}\mathcal{L}(t)\mathcal{P}=0,
\end{equation} 
but this depends
on the interplay of the initial state of the environment and the form of the interaction Hamiltonian. It is straightforward to show that for the Hamiltonian
under study in our paper, the condition for $\mathcal{P}\mathcal{L}(t)\mathcal{P}=0$ to be true is $\langle (S_{M}^{z}(t)) \rangle = 0$.
Since the initial state of the chain is the ground state of the chain Hamiltonian given by eq.~(1) in the main text, and it is known that the distribution of spin orientations for this state is symmetric, this expectation value is zero (the condition would also be fulfilled at any thermal equilibrium state of Hamiltonian (1)). 
This means that the second-order TCL equation for the qubit-chain system can be reduced to 
\begin{equation}
\label{tcleq2}
\frac{d}{dt} \mathcal{P}\rho_{SQ}(t) = \int_0^t ds \mathcal{P}\mathcal{L}(t)\mathcal{L}(s)\mathcal{P}\mathcal{P}\rho_{SQ}(t),
\end{equation}
which is equivalent to the Redfield equation \cite{redfield57}. It is important to note here that the equivalence is the result of the factorized initial conditions in our system
as well as the fulfillment of condition (\ref{plp}). Contrarily to the Redfield equation, the weak coupling limit does not need to be assumed here.

To find the second-order TCL equation, we insert the explicit form of the Liouville superoperator corresponding to the Hamiltonian in the interaction
picture (\ref{pict}). The derivation will yield $\frac{d}{dt} \langle i|\rho_{Q}(t)|i\rangle =0$,
meaning that the occupations of the qubit (the diagonal elements of the density matrix) do not evolve. This is in agreement with the fact that 
the Hamiltonian of the whole system fulfills the condition for qubit pure decoherence, $[H_Q,H_{SQ}]=0$. Thus, only the diagonal elements of the qubit
density matrix (the coherences) can evolve as a consequence of the qubit-chain interaction. 
This evolution (up to second-order corrections in TCL) is governed by the equation
\begin{equation}
	\frac{d}{dt} \rho_{01}(t)
	= -\rho_{01}(t) \frac{g^2}{2} \int_{-t}^t dt^{'} \langle S_{M}^{z}(t^{'})S_{M}^{z}(0)\rangle ,
\end{equation}
with $\rho_{01}(t)=\langle 0 | \rho_Q(t) | 1 \rangle $.
The formal solution to this equation is easily found and is explicitly given by eq.~(2) in the main text. 

\subsection{Transition at negative $\Delta$}

The transition between decaying and oscillatory behavior of the qubit decoherence occurs 
differently for positive and negative $\Delta$. For positive $\Delta$ it is gradual
and an admixture of higher order processes is visible in the decoherence, starting at $\Delta=1$,
while strictly oscillatory behavior sets in above $\Delta\approx 1.8$.
For negative $\Delta$ the transition occurs abruptly at $\Delta=1$. 

In Fig.~\ref{fig_SM1} we plot decoherence curves for $\Delta$ very close to $-1$, both above and 
below, in order to show that this is the case.

\begin{figure}[b]
\includegraphics[width=\linewidth]{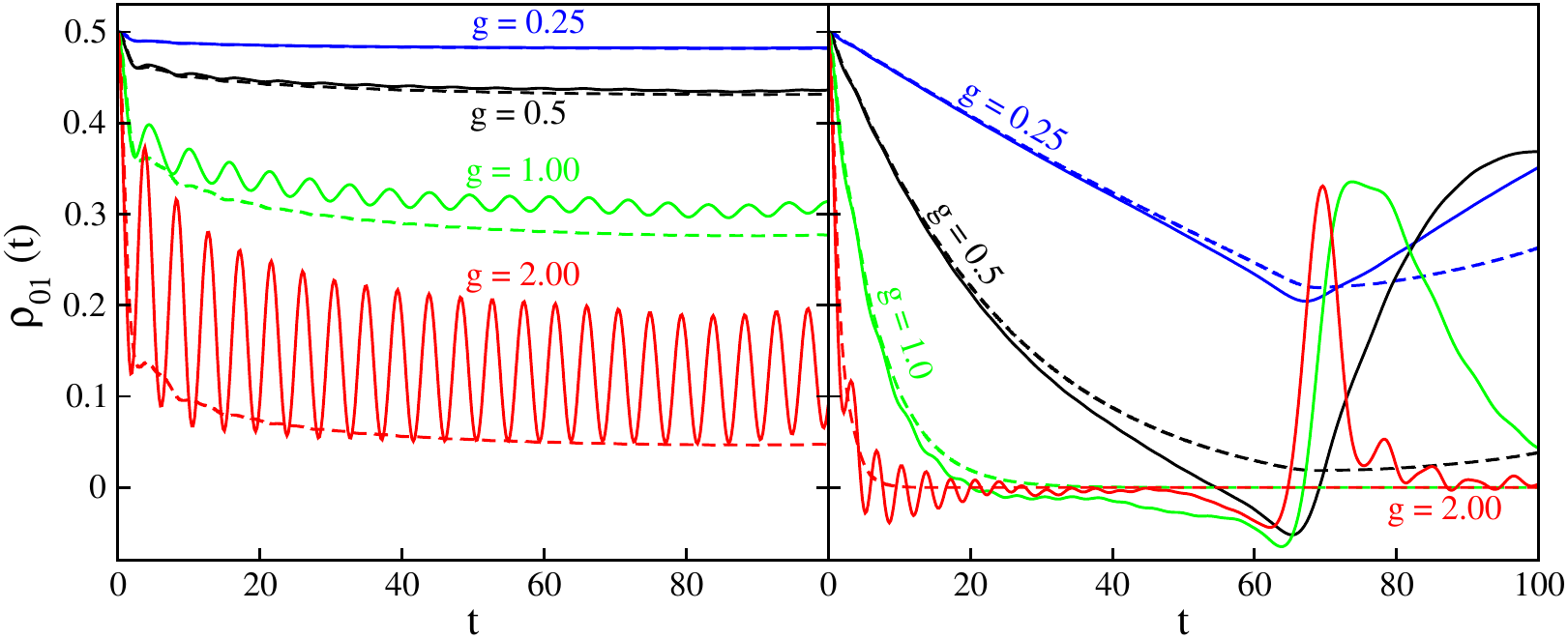}
\caption{Decoherence curves for set $\Delta = 0$ (left) and $\Delta = 1$ (right) for different values of the coupling strength (solid lines - TDVP, dashed lines - TCL).}
\label{fig_SM2}
\end{figure}

\subsection{Dependence on the coupling strength between the qubit and the chain}

The second-order TCL is a perturbative approach with respect to the qubit-chain interaction,
and as such it is assumed to work well only at small couplings between the two.
The strength of the interaction is quantified by the parameter $g$ in the Hamiltonian in the
main text. 

In the main text, the coupling is always set to $g=0.25 J$. In Fig.~\ref{fig_SM2}
we plot the TDVP and TCL curves at $\Delta = 0$ and $\Delta = 1$
(both in the range where the second-order TCL approximation holds for the results presented
in the main text) for a number of higher couplings. We observe that the correspondence between the TCL and TDVP results is surprisingly good even at large couplings, $g\ge J$, but it cannot account for the fast oscillatory behavior which is proportional to the coupling strength $g$ and is higher for the large anisotropy parameter
$\Delta = 1$. These oscillations are very prominent at large couplings, but TCL still provides a good approximation of the lower bound of the decoherence.

\end{document}